# Polarization Independent Ground State Optical Transitions in Closely Stacked InAs/GaAs Columnar Quantum Dots

Muhammad Usman
School of Physics, The University of Melbourne, Parkville, 3010, Victoria Australia
Email: usman@alumni.purdue.edu

**KEYWORDS:** Quantum Dots; Strain; Piezoelectric Potential; Degree of Polarization;

## 1 INTRODUCTION

In the last couple of decades, single and stacked InAs quantum dots (QDs) have been extensively studied for their potential application in the design of optical devices working at the telecommunication wavelengths (1300-1500 nm). Recent focus has been to achieve polarization insensitive optical transitions from InAs quantum dots by exploiting strain interactions between the QD layers in multi-layer quantum dot stacks. Experimental measurements [1, 2] have shown that the ground state optical transitions with isotropic polarization response can be achieved by closely stacking InAs quantum dots in the form of a 'columnar' quantum dot. This work analyses the polarization response of the columnar quantum dots consisting of up to 30 quantum dot layers. Our calculations indicate that a nearly isotropic polarization response can be achieved for about 11 quantum dots layers, which is in agreement to the recent experimental study where TM dominant optical spectra was measured for similar columnar QD stacks containing more than 9 layers [1].

## 2 METHODOLOGIES

The modelling and simulation of columnar quantum dots is a significant computational challenge. The anisotropy of charge density distribution $|\psi|^2$ inside the closely stacked quantum dots is significantly affected by the underlying crystal symmetry. The strain and piezoelectric potentials further lower the symmetry and needs to be incorporated at the atomistic level. The continuum modelling techniques such as k•p or effective mass approximation therefore fundamentally lack in sufficient physics to quantitatively model such devices. Furthermore, these very large size of quantum dot stacks (more than ten QD layers) require simulation domains containing of millions of atoms.

NEMO 3-D software package has been developed to handle large quantum dot systems with atomistic resolution [3]. It has shown the capabilities to simulate strain in large QD systems consisting of up to 64 million atoms [3] based on a valence force field (VFF) method [4]. The electronic structure calculations are performed by solving a twenty band $sp^3d^5s^*$ empirical tight binding (TB) Hamiltonian [5]. The linear and quadratic piezoelectric potentials are included in the calculations. The interband optical transitions are computed using Fermi's golden rule by the squared magnitude of the optical matrix elements summed over degenerate electronic states [6,7].

## 3 RESULTS AND DISCUSSIONS

Figure 1 (a) shows a schematic diagram of the simulated system. An InAs columnar QD consisting of 'N' ($1 \leq N \leq 30$) quantum dot layers is placed inside a GaAs buffer. The columnar QD consists of 'N' 1.5 nm tall InAs QDs placed on top of 1 monolayer (ML) thick InAs wetting layers. The base diameter 'B' of each QD is 17 nm and the height 'H' is N*1.5 nm. The size of the GaAs buffer (60x60x100 nm$^3$, containing around 23 million atoms) is chosen to be large enough to properly accommodate the long-range effects of strain and piezoelectric fields. The atomistic relaxation is performed with realistic boundary conditions *i.e.* the bottom fixed, the lateral dimensions periodic, and the top free to relax. The electronic structure calculations have closed boundary conditions in all directions.

Figure 1 (b) plots the ground state optical transition wavelength (E1-H1) as a function of the number of QD layers N. As N increases, the wavelength also increases, until it saturates for N>10. Similar results were shown in the experimental study [1]. The saturation of the wavelength at ~1550 nm is due to the change in the orientation of the electronic states from 2-D (*xy*) symmetry to 1-D (*z*) symmetry inside the columnar QD (see figure 2).

The polarization response of QD samples is characterized in terms of degree of polarization (DOP) [1, 2] of the optical transitions. The DOP is defined as the ratio (TE-TM)/(TE+TM) [1]. Figure 1 (c) presents the plots of the DOP and the aspect ratio (AR=H/B) as function of N. For relatively flat QDs (AR<0.6), the (001) confinement is very strong. The topmost valence band state is dominantly a heavy hole (HH) like state with a weak light hole (LH) contribution; therefore, the TE mode coupling is much stronger than the TM mode coupling. As the number quantum dot layers N increases, the (001) confinement is relaxed. For large values of the AR, the in-plane strain becomes strong and hence the electronic states start becoming more confined along the *x-y* directions. The reversal of biaxial strain sign increases the LH contribution and as a result, the TE mode decreases and the TM mode increases [8]. For N~11, the magnitudes of the TE and TM modes become nearly equal (TE/TM ~ 1.07) and as a result, the DOP is ~ 0.036. Further increase in N results in the TM mode becoming stronger than the TE mode and hence the DOP takes up negative values.

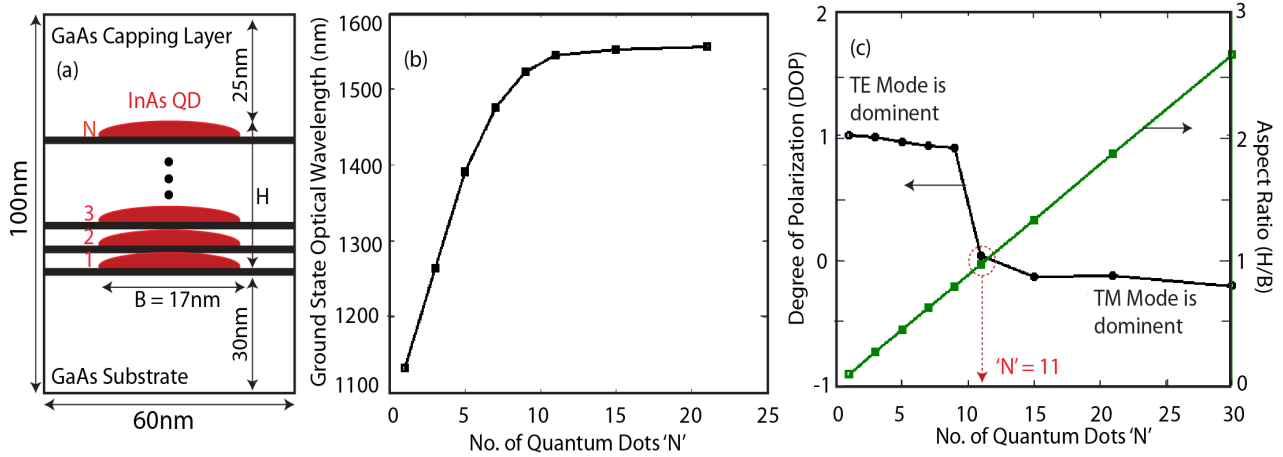

**Figure 1:** (a) Schematic of the simulated system. (b) Plot of ground state optical wave length (E1-H1) as a function of the number of quantum dot layers N. (c) Plot of the degree of polarization (DOP) and aspect ratio (H/B) as a function of the number of quantum dot layers N.

Figure 2 shows the wave function plots of the highest valence band state H1 and the lowest conduction band state E1 for N=1, N=11, and N=21. For N=1, the strong (001) confinement results in nearly flat wave functions in the *xy* plane. For N=11 and N=21, the elongation of the wave functions along the (001) direction is quite evident.

In summary, our calculations show that the desirable isotropic polarization response (DOP~0) can be achieved from the columnar InAs QDs for N~11. The computed results are in good agreement with the reported measurements. The demonstration of the isotropic polarization response of the ground state optical transition at ~1550 nm wavelength clearly shows a promising potential of such columnar QDs for the design of optical devices such as semiconductor optical amplifiers (SOAs).

## 4 ACKNOWLEDGEMENTS

The computational resources were provided by National Science Foundation (NSF) funded Network for Computational Nanotechnology (NCN) through nanoHUB.org.

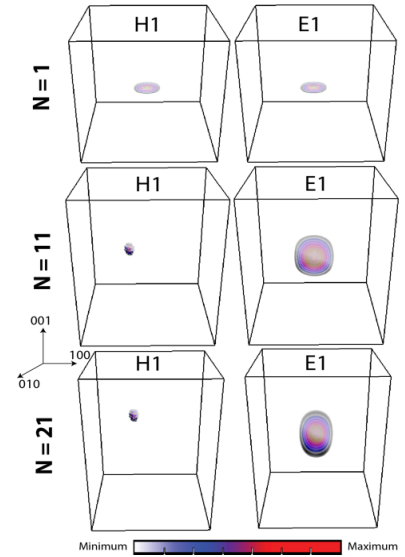

**Figure 2:** Plots of the wave functions for the highest valence band state H1 and the lowest conduction band state E1 are shown for N=1, N=11, and N=21.